\documentclass[a4paper, 11pt]{amsart}
\usepackage{graphicx}
\usepackage{amsfonts}
\usepackage{amsmath}
\usepackage{amssymb}

\begin{document}

\title{Magic informationally complete POVMs\\ with permutations
}

\author{Michel Planat$\dag$ and Zafer Gedik$\ddag$}
\address{$\dag$Universit\'e de Bourgogne/Franche-Comt\'e, Institut FEMTO-ST CNRS UMR 6174, 15 B Avenue des Montboucons, F-25044 Besan\c con, France.}
\email{michel.planat@femto-st.fr}

\address{$\ddag$Faculty of Engineering and Natural Sciences, Sabanci University, Tuzla, Istanbul, 34956, Turkey.}
\email{gedik@sabanciuniv.edu}


\begin{abstract}

Eigenstates of permutation gates are either stabilizer states (for gates in the Pauli group) or magic states, thus allowing universal quantum computation [M. Planat and Rukhsan-Ul-Haq, Preprint 1701.06443]. We show in this paper that a subset of such magic states, when acting on the generalized Pauli group, define (asymmetric) informationally complete POVMs. Such IC-POVMs, investigated in dimensions $2$ to $12$, exhibit simple finite geometries in their projector products and, for dimensions $4$ and $8$ and $9$, relate to two-qubit, three-qubit and two-qutrit contextuality.  

\end{abstract}

\maketitle

\vspace*{-.5cm}

\normalsize

\section{Introduction}

Sometimes a field of knowledge gets enriched just by looking at it on a different perspective. Here we are interested in informationally complete (IC) measurements on an unknown density matrix $\rho$ with the perspective of universal quantum computing. In the former subfield, one knows how to build group covariant symmetric measurements (SIC-POVMs) that follow from the action of the generalized Pauli group $\mathcal{P}_d$ on a well choosen \lq fiducial' state \cite{Renes2004,Flammia2006,Appleby2005}. In the latter subfield, the group $\mathcal{P}_d$ needs to be extended by a well choosen \lq magic' state of the corresponding dimension to allow universal quantum computation \cite{Bravyi2004,Reichardt2005}. S. Bravyi and A. Kitaev \cite{Bravyi2004} introduced the principle of ‘magic state distillation’: universal quantum computation may be realized thanks to the stabilizer formalism (Clifford group unitaries, preparations and measurements) and the ability to prepare an ancilla in an appropriate single qubit mixed state. Following \cite[Sec. IIC]{Veitch2014}, in this paper, a non-stabilizer pure state will be called a magic state. When is such a \lq magic' state \lq fiducial' for an IC-POVM? To address this question, we restrict our choice to eigenstates of permutation gates not living in $\mathcal{P}_d$ (the stabilizer subgroup of unitaries) as in the recent paper \cite{Planat2017}. We recover the Hesse SIC for $d=3$ and discover asymmetric IC-POVMs for $d>3$.

In this paper we first remind in Sec. 1 a few necessary concepts for our purpose: POVM concepts and the generalized Pauli group. In Sec. 2, we apply the methodology to the derivation of IC-POVM's in dimensions $2$ to $12$, then we establish the link to some finite geometries and to two-qubit, three-qubit and two-qutrit contextuality.  Sec. 3 summarizes the results. 

A POVM is a collection of positive semi-definite operators $\{E_1,\ldots,E_m\}$ that sum to the identity. In the measurement of a state $\rho$, the $i$-th outcome is obtained with a probability given by the Born rule $p(i)=\mbox{tr}(\rho E_i)$. For a minimal IC-POVM, one needs $d^2$ one-dimensional projectors $\Pi_i=\left|\psi_i\right\rangle \left\langle \psi_i \right|$, with $\Pi_i=d E_i$, such that the rank of the Gram matrix with elements $\mbox{tr}(\Pi_i\Pi_j)$, is precisely $d^2$.

A SIC-POVM obeys the remarkable relation \cite{Renes2004}

$$\left |\left\langle \psi_i|\psi_j \right \rangle \right |^2=\mbox{tr}(\Pi_i\Pi_j)=\frac{d\delta_{ij}+1}{d+1},$$

that allows the recovery of the density matrix as \cite{Fuchs2004}

$$\rho=\sum_{i=1}^{d^2}\left[ (d+1)p(i)-\frac{1}{d} \right]\Pi_i.$$

This type of quantum tomography is often known as quantum-Bayesian, where the $p(i)$'s represent agent's Bayesian degrees of belief, because the measurement depends on the filtering of $\rho$ by the selected SIC (for an unknown classical signal, this looks similar to the frequency spectrum).

In this paper, we discover new IC-POVMs (i.e. whose rank of the Gram matrix is $d^2$) and with Hermitian angles $\left |\left\langle \psi_i|\psi_j \right \rangle \right |_{i \ne j} \in A=\{a_1,\ldots,a_l\}$, a discrete set of values of small cardinality $l$. A SIC is equiangular with $|A|=1$ and $a_1=\frac{1}{\sqrt{d+1}}$.

The states encountered below are considered to live in a cyclotomic field $\mathbb{F}=\mathbb{Q}[\exp(\frac{2i\pi}{n})]$,
 with $n=\mbox{GCD}(d,r)$, the greatest common divisor of $d$ and $r$, for some $r$. 
The Hermitian angle is defined as $\left |\left\langle \psi_i|\psi_j \right \rangle \right |_{i \ne j}=\left\|(\psi_i,\psi_j)\right\|^{\frac{1}{\mbox{\footnotesize deg}}}$, where $\left\|.\right\|$ means the field norm \cite[p. 162]{Cohen1996} of the pair $(\psi_i,\psi_j)$ in $\mathbb{F}$ and $\mbox{deg}$ is the degree of the extension $\mathbb{F}$ over the rational field $\mathbb{Q}$. For the IC-POVMs under consideration below, in dimensions $d=3$, $4$, $5$, $6$ and $7$, one has to choose $n=3$, $12$, $20$, $6$ and $21$ respectively, in order to be able to compute the action of the Pauli group. Calculations are performed with Magma.

\subsection*{The single qubit SIC-POVM}

To introduce our methodology, let us start with the qubit magic state

$$\left| T\right\rangle=\cos(\beta)\left| 0\right\rangle + \exp{(\frac{i\pi}{4})}\sin(\beta)\left| 1\right\rangle,~~ \cos(2\beta)=\frac{1}{\sqrt{3}},$$
employed for universal quantum computation \cite{Bravyi2004}. It is defined as the $\omega_3=\exp(\frac{2i\pi}{3})$-eigenstate of the $SH$ matrix [the product of the Hadamard matrix $H$ and the phase gate $S=\bigl( \begin{smallmatrix} 
  1 & 0\\
  0 & i 
\end{smallmatrix}\bigr)$].

Taking the action on $\left| T\right\rangle$  of the four Pauli gates $I$, $X$, $Z$ and $Y$, the corresponding (pure) projectors  $\Pi_i=\left|\psi_i\right\rangle \left\langle \psi_i \right|,i=1\ldots 4$, sum to twice the identity matrix thus building a POVM and the pairwise distinct products satisfy $\left|\left \langle \psi_i \right|\psi_j \right\rangle|^2=\frac{1}{3}$. The four elements $\Pi_i$ form the well known $2$-dimensional SIC-POVM \cite[Sec. 2]{Renes2004}.

In contrast, there is no POVM attached to the magic state $\left| H\right\rangle=\cos(\frac{\pi}{8})\left| 0\right\rangle + \sin(\frac{\pi}{8})\left| 1\right\rangle$.

\subsection*{The generalized Pauli group}

Later, we construct IC-POVMs using the covariance with respect to the generalized Pauli group.
Let $d$ be a prime number, the qudit Pauli group is generated by the shift and clock operators as follows
\begin{eqnarray}
&X\left|j \right\rangle = \left|j+1 \mod d \right\rangle \nonumber \\
&Z \left |j \right\rangle=\omega^j \left|j \right\rangle 
\label{eqn1}
\end{eqnarray}
with $\omega=\exp(2i\pi/d)$ a $d$-th root of unity. In dimension $d=2$, $X$ and $Z$ are the Pauli spin matrices $\sigma_x$ and $\sigma_z$.

A general Pauli (also called Heisenberg-Weyl) operator is of the form

\begin{equation}
T_{(m,j)} = \left\{
    \begin{array}{ll}
        i^{jm}Z^m X^j & \mbox{if } d=2 \\
        \omega^{-jm/2}Z^m X^j & \mbox{if } d \ne 2.
    \end{array}
\right.
\label{eqn2}
\end{equation}
where $(j,m) \in \mathbb{Z}_d \times \mathbb{Z}_d$.
For $N$ particules, one takes the Kronecker product of qudit elements $N$ times.

Stabilizer states are defined as eigenstates of the Pauli group.

\section{Permutation gates, magic states and informationally complete measurements}

In the approach of magic states through permutation groups, dimension two is trivial since the symmetric group $S_2$ only contains the identity $I=(1,2)$ and the shift gate $X=(2,1)\equiv \bigl( \begin{smallmatrix} 
  0 & 1\\
  1 & 0 
\end{smallmatrix}\bigr)$, that live in the ordinary Pauli group $\mathcal{P}_2$. No magic state may be derived from two-dimensional permutation groups.

The situation changes as soon as $d \ge 3$ with a wealth of magic states \cite{Planat2017} having a potential usefulness for our purpose of defining IC-POVMs. From now we focus on magic groups generated by two magic permutation gates.

\subsection*{In dimension three}

The symmetric group $S_3$ contains the permutation matrices $I$, $X$ and $X^2$ of the Pauli group, where $X=\bigl( \begin{smallmatrix} 
  0 & 1&0\\
  0&0 & 1 \\
	1&0&0
\end{smallmatrix} \bigr)\equiv (2,3,1)$ and three extra permutations $\bigl( \begin{smallmatrix} 
  1 & 0&0\\
  0&0 & 1 \\
	0&1&0
\end{smallmatrix} \bigr)\equiv (2,3)$, $\bigl( \begin{smallmatrix} 
  0 & 0&1\\
  0&1 & 0 \\
	1&0&0
\end{smallmatrix} \bigr)\equiv (1,3)$ and $\bigl( \begin{smallmatrix} 
  0 & 1&0\\
  1&0 & 0 \\
	0&0&1
\end{smallmatrix} \bigr)\equiv (1,2)$, that do not lie in the Pauli group but are parts of the so-called Clifford group (the normalizer of the Pauli group in the unitary group).

\begin{figure}[ht]
\includegraphics[width=6cm]{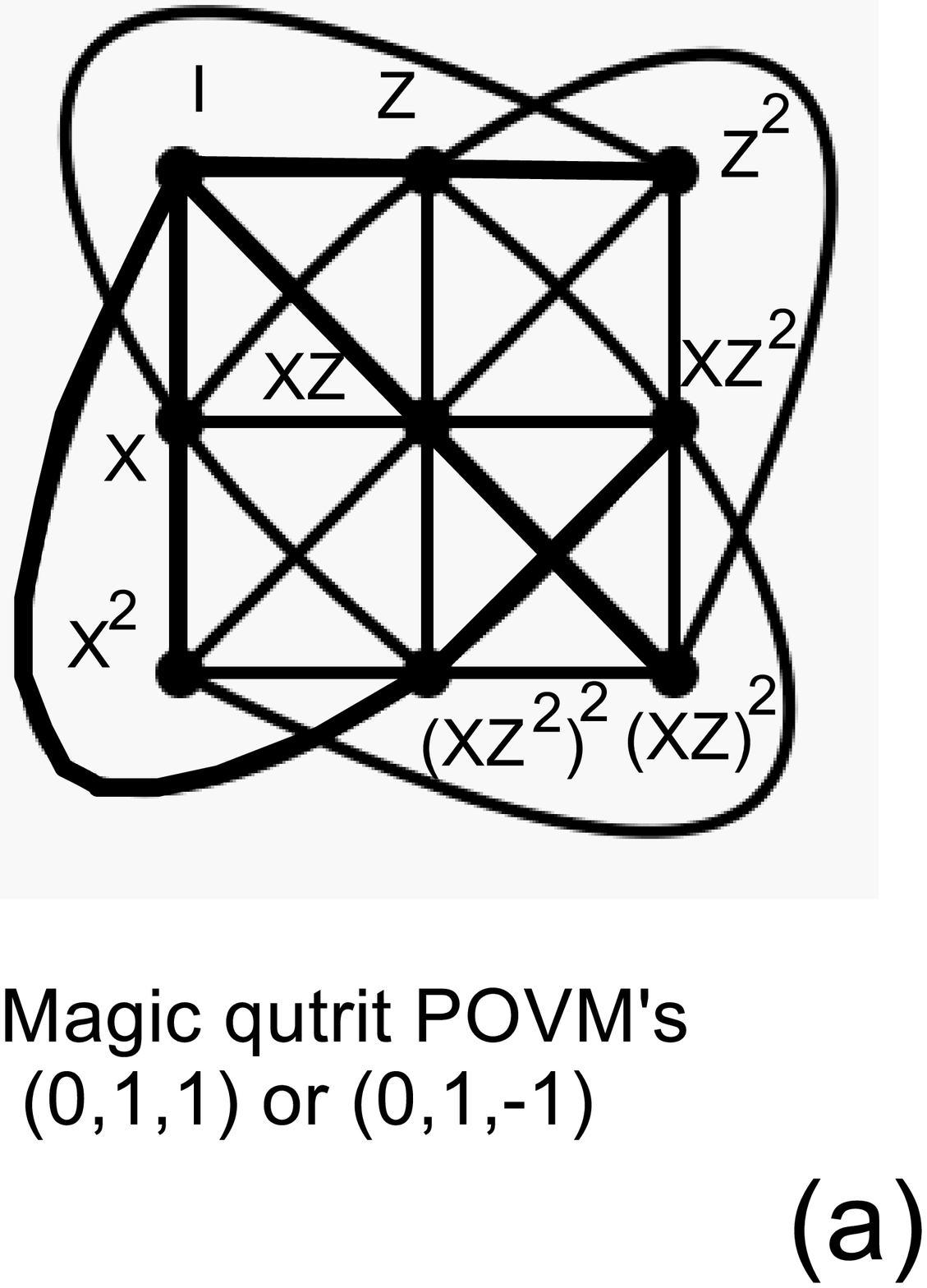}
\includegraphics[width=6cm]{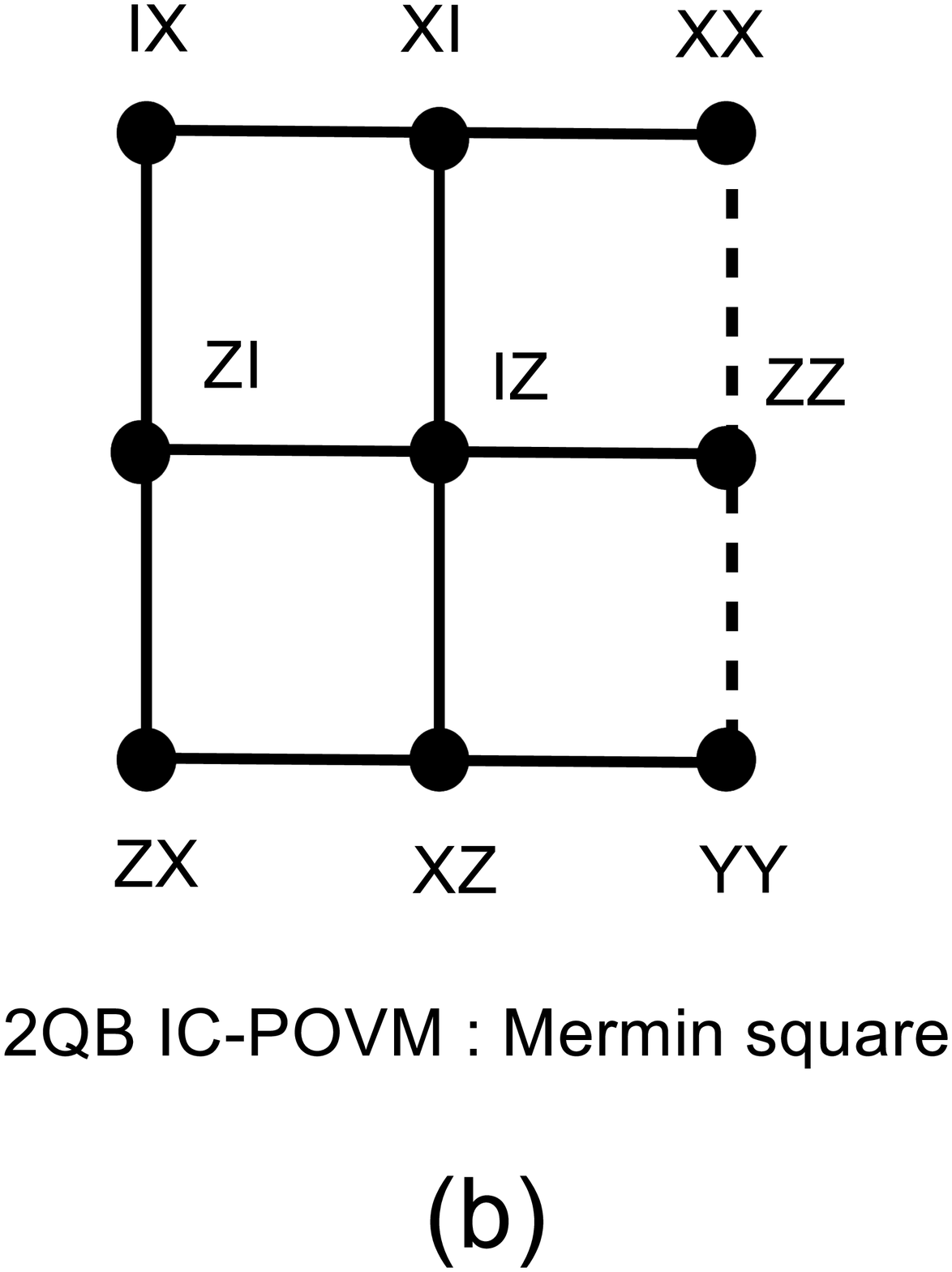}
\caption{ (a) The Hesse configuration resulting from the qutrit POVM. The lines of the configuration correspond to traces of triple products of the corresponding projectors equal to $\frac{1}{8}$ [for the state $(0,1,-1)$] and $\pm \frac{1}{8}$ [for the state $(0,1,1)$]. The configuration is labelled in terms of the qutrit operators acting on the magic state. Bold lines feature the lines where all operator pairs are commuting. 
(b) The triple products  of the four dimensional IC-POVM whose trace equal $\pm \frac{1}{27}$ and simultaneously equal plus or minus the identity matrix $\mathcal{I}$ ($-\mathcal{I}$ for the dotted line). This picture identifies to the well known Mermin square which allows a proof of the Kochen-Specker theorem.
 }
\end{figure}

Taking the eigensystem of the latter matrices, it is not difficult check that there exists two types of qutrit magic states of the form $(0,1,\pm1) \equiv \frac{1}{\sqrt{2}} (\left| 1\right\rangle\pm \left| 2\right\rangle)$. Then, taking the action of the nine qutrit Pauli matrices, one arrives at the well known Hesse SIC \cite{Bengtsson2010,Tabia2013, Hughston2007}.

The Hesse configuration shown in Fig. 1a is a configuration $[9_4,12_3]$ with $9$ points and $12$ lines, $4$ lines incident on every point and $3$ points on a line. It can also be seen as the $3$-dimensional affine plane. The reason it occurs in the context of the $3$-dimensional SIC is as follows. The SIC relations are $\mbox{tr}(\Pi_i \Pi_j)_{i \ne j}=\frac{1}{4}$ and, if one takes all projectors satisfying the triple product relation $\mbox{tr}(\Pi_i \Pi_j \Pi_k)_{i \ne j \ne k}=\pm \frac{1}{8}$, the corresponding triples $(i,j,k)$ define the Hesse configuration. For the Hesse SIC built from the magic state $(0,1,-1)$, one only needs the plus sign to recover the Hesse geometry, but for the Hesse SIC built from the magic state $(0,1,1)$ both signs are needed (see also \cite{Tabia2013}). 

Observe that the configuration in Fig. 1a is labeled in terms of qutrit operators acting on the magic states instead of the projector themselves.

\subsection*{In dimension four}

In dimension four and higher, the strategy is to restrict to permutation groups whose two generators are magic gates, gates showing one entry of $1$ on their main diagonals. From now we call such a group a magic group. This only happens for a group isomorphic to the alternating group $$A_4\cong \left\langle   \left(\begin{smallmatrix} 
  1 & 0&0&0\\
  0&0 &0& 1 \\
	0&1&0&0\\
	0&0&1&0
\end{smallmatrix} \right),
 \left( \begin{smallmatrix} 
  0 & 1&0&0\\
  0&0 &1& 0 \\
	1&0&0&0\\
	0&0&0&1
\end{smallmatrix} \right) \right\rangle.$$ One finds magic states of type $(0,1,1,1)$ and $(0,1,-\omega_6,\omega_6-1)$, with $\omega_6=\exp(\frac{2i\pi}{6})$.
 \cite[Sec. 3.3]{Planat2017}.

Taking the action of the two-qubit Pauli group on the latter type of state, the corresponding pure projectors sum to four times the identity (to form a POVM) and are independent, with the pairwise distinct products satisfying the dichotomic relation  $\mbox{tr}(\Pi_i \Pi_j)_{i \ne j}=\left|\left \langle \psi_i \right|\psi_j \right\rangle|_{i \ne j}^2
 \in \{\frac{1}{3},\frac{1}{3^2}\}$. Thus the $16$ projectors $\Pi_i$ build an asymmetric informationally complete measurement not discovered so far.

The organization of triple products of projectors whose trace is $\pm \frac{1}{27}$ and simultaneously equal plus or minus the identity matrix $\mathcal{I}$ is shown in Fig. 1b. Instead of labelling coordinates as projectors one may label them with the two-qubit operators acting on the magic state. As a result, the two-qubit $(3 \times 3)$-grid identifies to the standard Mermin square that is known to allow an operator proof of the Kochen-Specker theorem \cite{Mermin1993,Planat2012}.

\subsection*{In dimension five}

Still restricting to permutation groups generated by two magic gates (magic groups), the smallest group is isomorphic to the semidirect product $\mathbb{Z}_5 \rtimes \mathbb{Z}_4$ of cyclic groups $\mathbb{Z}_4$ and $\mathbb{Z}_5$ \cite[Sec 3.4]{Planat2017}. One finds magic states of type $(0,1, 1,1,1)$, $(0,1,-1,-1,1)$ and $(0,1,i,-i,-1)$. The latter two types allow to construct IC-POVM's such that the pairwise distinct products satisfy  $\left|\left \langle \psi_i \right|\psi_j \right\rangle|^2=\frac{1}{4^2}$, that is the POVM is equiangular with respect to the field norm defined in the introduction. The first type of magic state is dichotomic with values of the products $\frac{1}{4^2}$ and $(\frac{3}{4})^2$. The trace of pairwise products of (distinct) projectors is not constant. For example, with the state $(0,1,-1,-1,1)$, one gets a field norm equiangular IC-POVM in which the trace is trivalued: it is either $1/16$ or $(7\pm 3 \sqrt{5})/32$. For the state $(0,1,i,-i,-1)$, there are five values of the trace.

 With the symmetric group $S_5$, one builds magic states of type $(0,0,1,1,1)$ and IC-POVM's with dichotomic values of the distinct pairwise products equal to $(\frac{1}{3})^2$ and $(\frac{2}{3})^2$.

Let us concentrate on the equiangular POVM.
Traces of triple products with constant value $-\frac{1}{4^3}$ define lines organized into a geometric configuration of type $(25_{12},100_3)$. Lines of the configuration have one or two points in common. The two-point intersection graph consists of $10$ disjoint copies of the Petersen graph. One such a Petersen graph is shown in Fig. 2a, the vertices of the graph correspond to the lines and the edges correspond  to the one-point intersection of two lines. As before the labelling is in terms of the operators acting on the magic state.

Similar Petersen graphs occur in the organization of triple products for the other five-dimensional IC-POVMs.

\begin{figure}[ht]
\includegraphics[width=6cm]{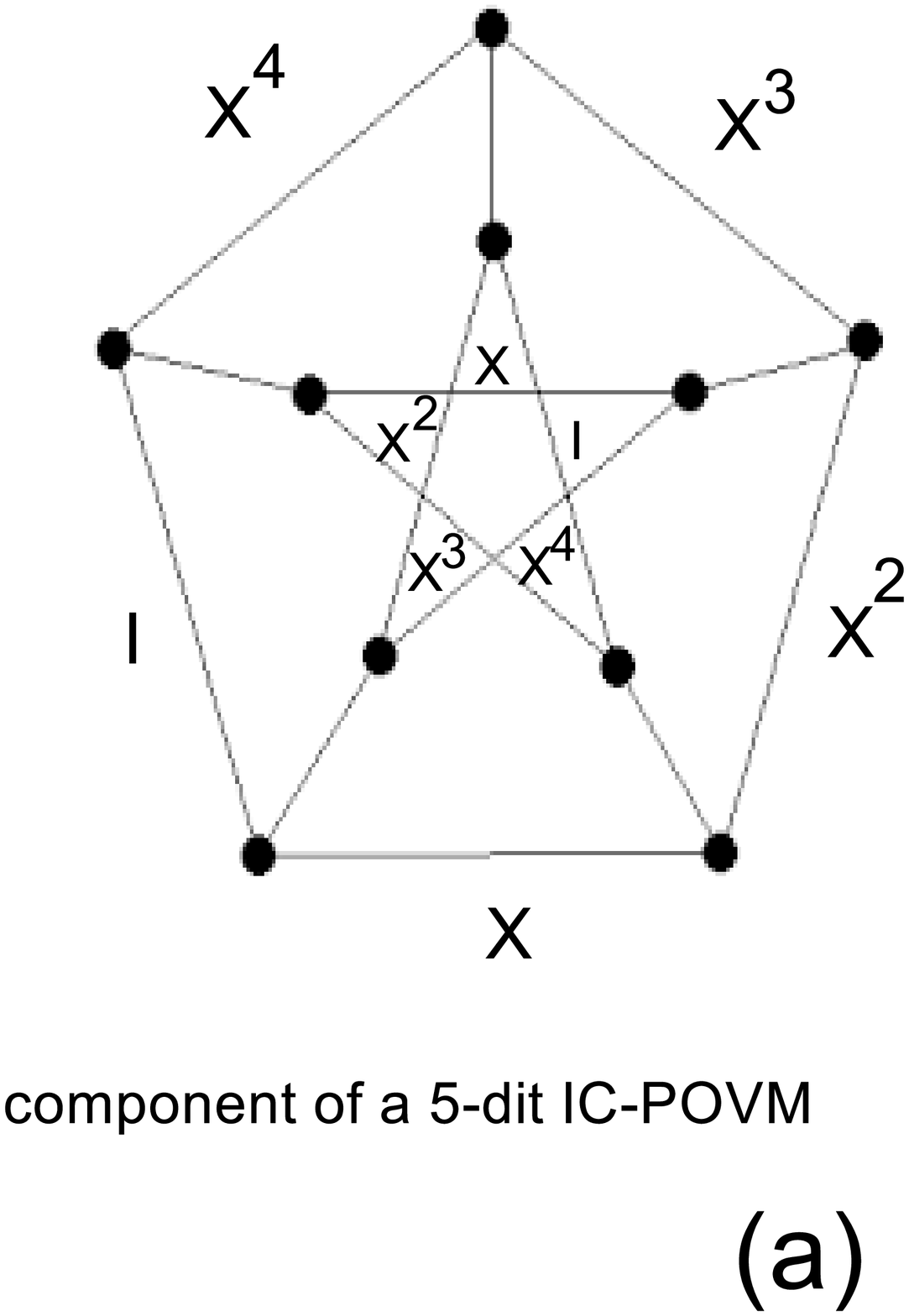}
\includegraphics[width=5cm]{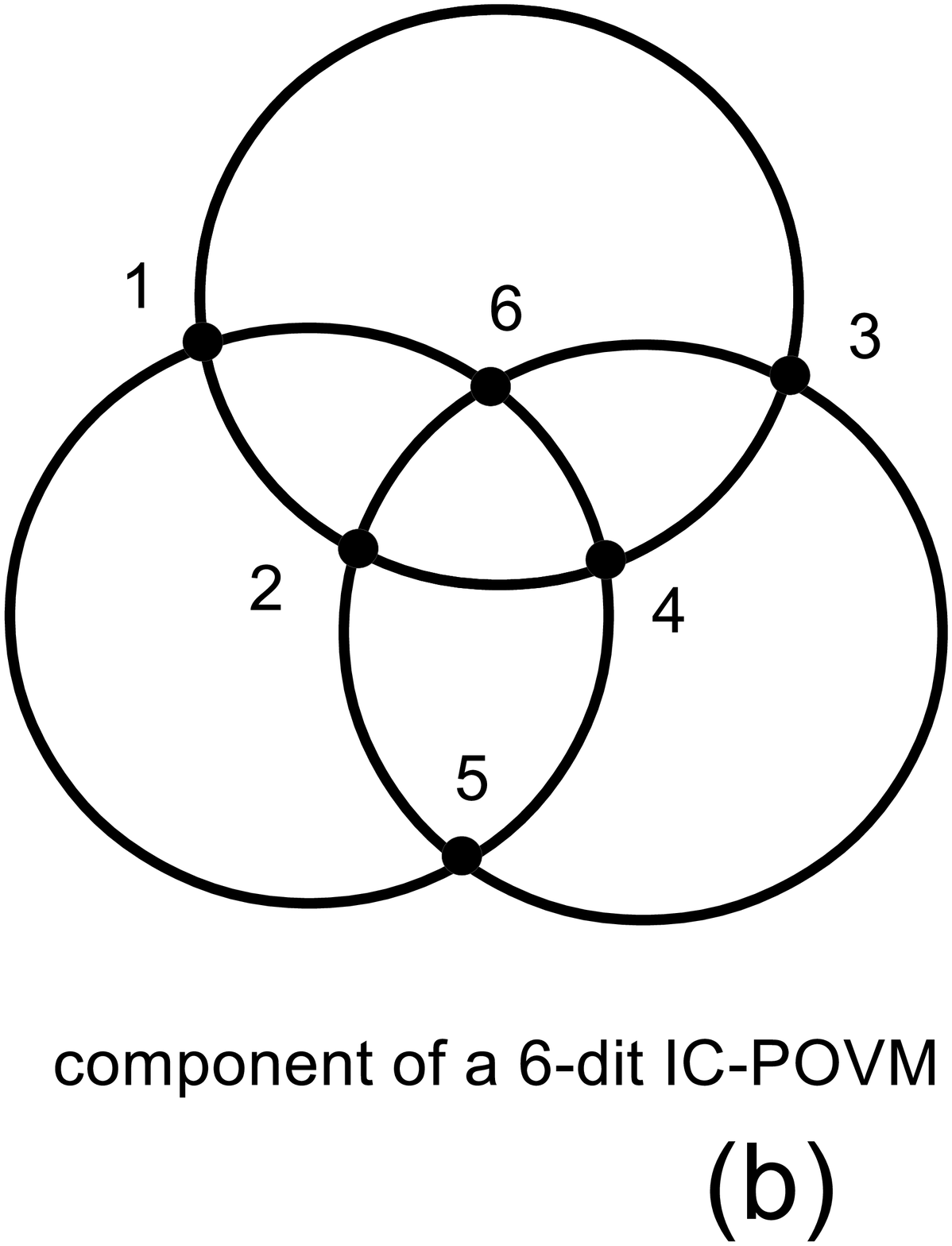}
\caption{(a) A one-point intersection graph for the lines of the $5$-dit equiangular IC-POVM defined from the triple products  of constant trace $-\frac{1}{4^3}$. (b) A component of the $6$-dit IC-POVM with magic state $(0,1,\omega_6-1,0,-\omega_6,0)$ through the action of Pauli operators $1$ to $6$: the lines correspond  to $4$-tuples products of projectors with constant trace $\frac{1}{9}$ and simultaneously of products equal to $\pm\mathcal{I}$.  There are two disjoint copies looking like Borromean rings with points as \newline $[1\ldots 6]=[\mathcal{I},ZX^3,Z^2,Z^3X^3,Z^4,Z^5X^3]$ and $[1\ldots 6]=[X^4,Z,Z^2X^3,Z^3,Z^4X^3,Z^5]$.
 }
\end{figure}

\subsection*{In dimension six}

With the alternating group $A_6$ generated by two magic gates, one finds an IC-POVM
associated to a magic state such as $(0,1,\omega_6-1,0,-\omega_6,0)$ with  $\mbox{tr}(\Pi_i \Pi_j)_{i \ne j}=\left|\left \langle \psi_i \right|\psi_j \right\rangle|^2_{i \ne j}=\frac{1}{3}\mbox{or}~~\frac{1}{3^2}$.

Taking the trace of $4$-tuple products of projectors whose value is $\frac{1}{9}$ and simultaneously equal $\pm \mathcal{I}$, one gets two copies of a geometry looking like a Borromean ring as shown in Fig. 2b. 

\subsection*{In dimension seven}

Using a magic group isomorphic to $\mathbb{Z}_7 \rtimes \mathbb{Z}_6$ and the magic state $(1,-\omega_3-1,-\omega_3,\omega_3,\omega_3+1,-1,0)$,
 one arrives at an equiangular IC-POVM satisfying $\left|\left \langle \psi_i \right|\psi_j \right\rangle|^2_{i \ne j}=\frac{1}{6^{2}}$. Other magic states are also found that define IC-POVM's with dichotomic products. But no simple structure of the higher order products has been found.

\subsection*{In dimension eight}

In dimension $d=8$, no IC-POVM was discovered from permutation groups. But it is time to introduce the well known Hoggar SIC \cite{Zhu2015,Stacey2016}. The Hoggar SIC follows from the action of the three-qubit Pauli group on a fiducial state such as $(-1  \pm i, 1,1,1,1,1,1,1)$.

\begin{figure}[ht]
\includegraphics[width=8cm]{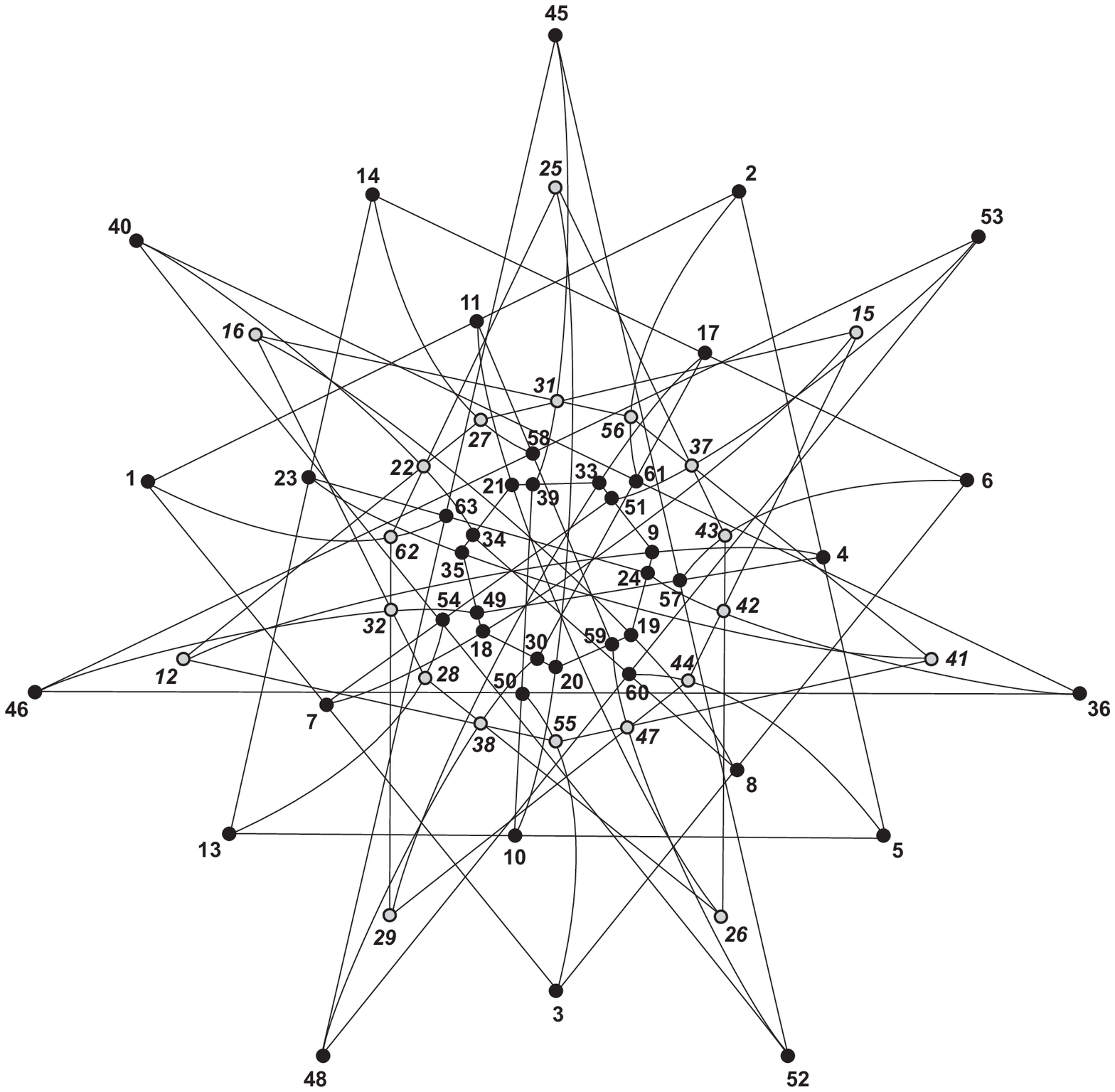}
\caption{The dual of the generalized hexagon $GH(2,2)$. Grey points have the structure of an embedded generalized hexagon $GH(2,1)$ \cite{Frohardt1994}.
 }
\end{figure}

\begin{figure}[ht]
\includegraphics[width=5cm]{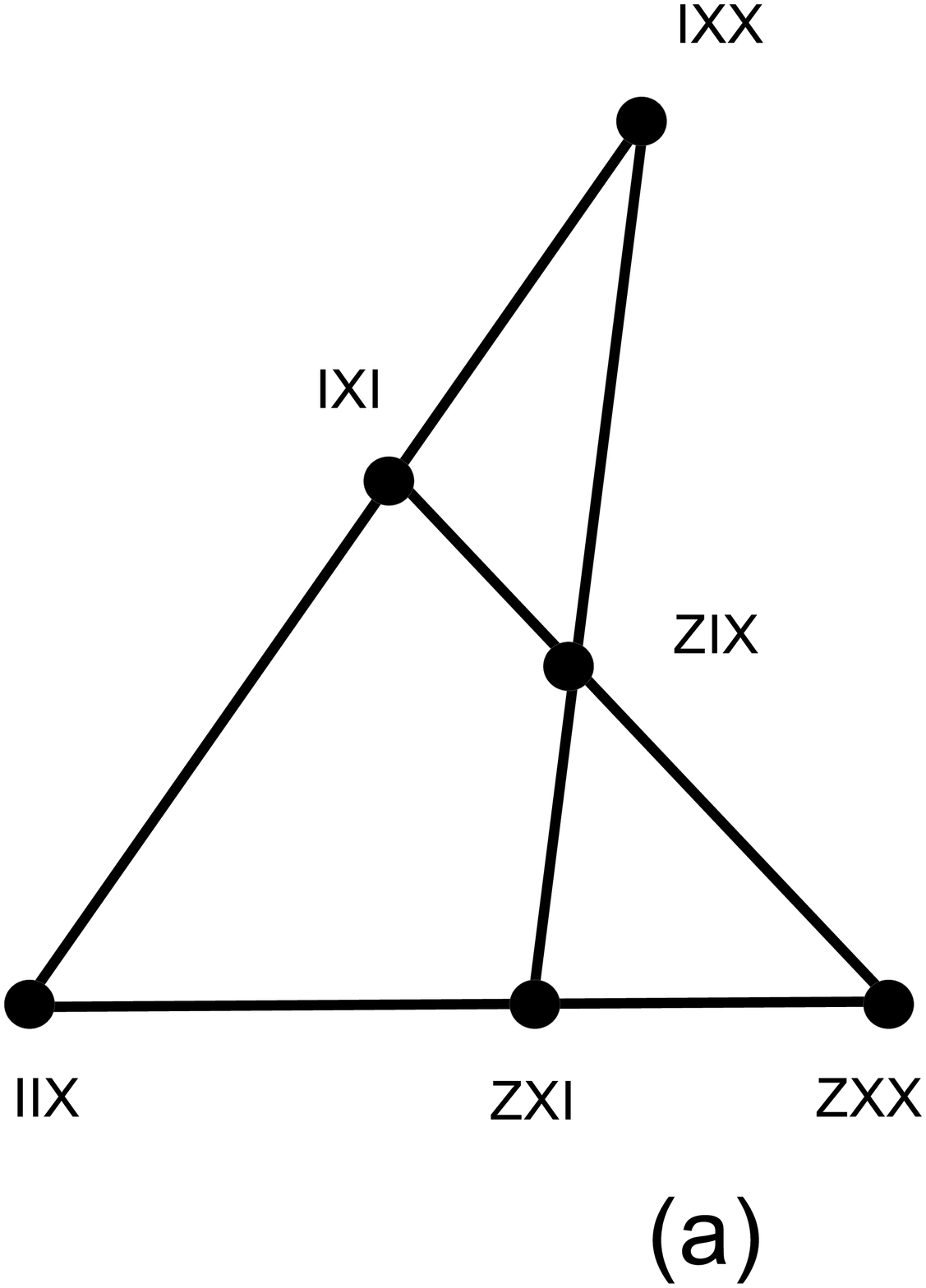}
\includegraphics[width=5cm]{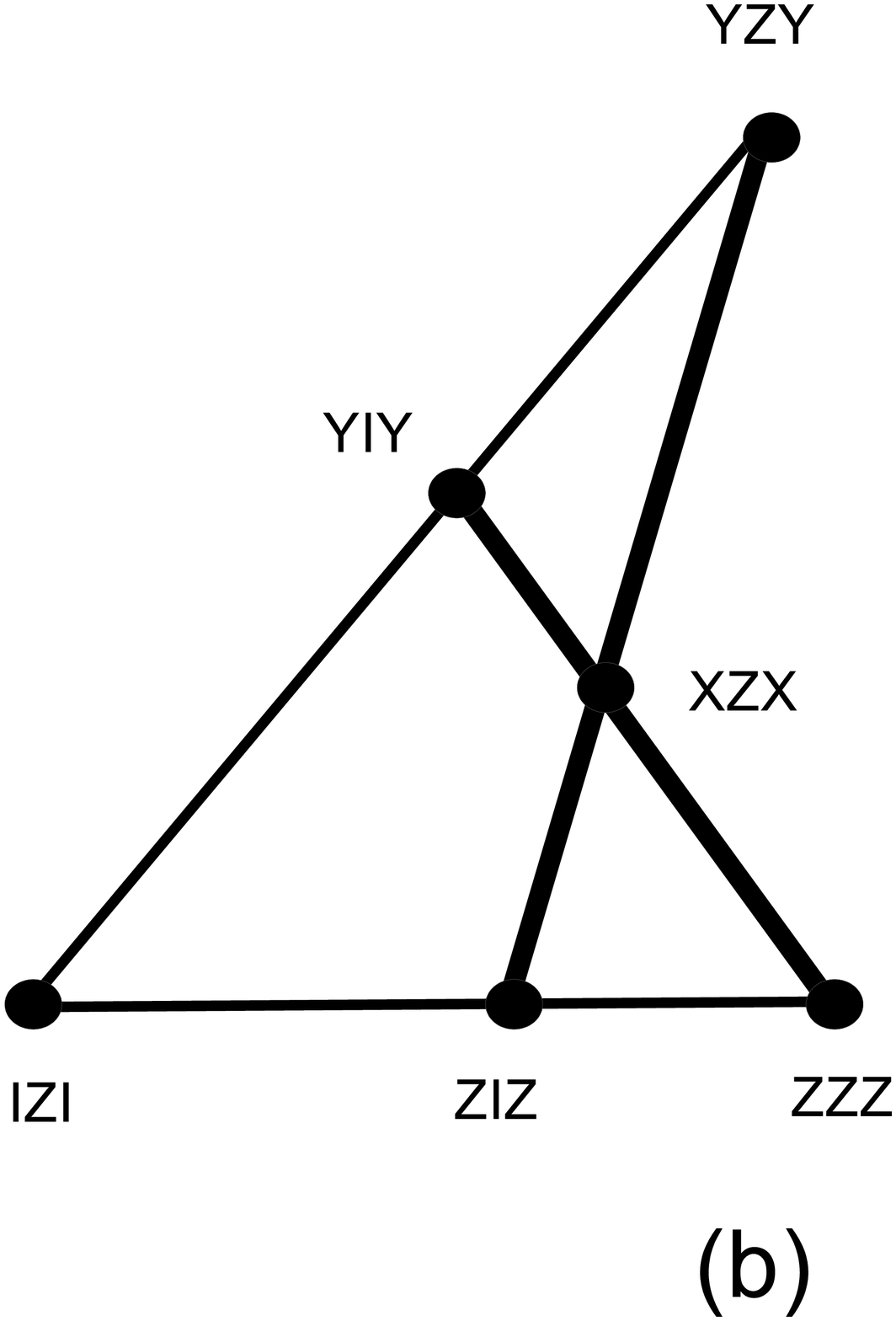}
\caption{ Two types of Pasch blocks in the structure of trace $\frac{1}{27}$ triple products of a Hoggar SIC.  Thin (resp thick) lines are for triple products equal to $\mathcal{I}$ (resp. $-\mathcal{I}$).
 }
\end{figure}

 It has been found that triple products are related to combinatorial designs \cite{Stacey2016}. There are $4032$ (resp. $16128$) triples of projectors whose products have trace equal to $-\frac{1}{27}$ (resp. $\frac{1}{27}$) \cite[(29)]{Stacey2016}. Within the $4032$ triples, those whose product of projectors equal $\pm \mathcal{I}$ (with $\mathcal{I}$ the identity matrix) are organized into a geometric configuration $[63_3]$ whose incidence graph is of spectrum $[6^1,3^{31},-1^{27},-3^{14}]$ and automorphism group $G_2(2)=U_3(3) \rtimes \mathbb{Z}_2$ of order $12096$, as in \cite{Planat2013}. It is known that there exists two isospectral configurations of this type, one is the so-called generalized hexagon $GH(2,2)$ (also called split Cayley hexagon) and the other one is its dual \cite{Frohardt1994}. These configurations are related to the $12096$ Mermin pentagrams that build a proof of the three-qubit Kochen-Specker theorem \cite{Planat2013,Levay2017}. From the structure of hyperplanes of our $[63_3]$ configuration, one learns that we are concerned with the dual of $G_2$ as shown in Fig. 3 (see also \cite[Fig. 6a]{Planat2016}).

Similarly within the $16128$ triples, set of projectors whose triples equal $\pm \mathcal{I}$ are organized into a configuration $[63_{12},252_3]$ whose incidence graph has spectrum $[33^1,15^{14},9^{21},5^{27},-3^{189}]$ and automorphism group $G_2(2)$ again. The graph shows $63$ maximum cliques of size $4$ and $72$ of size $7$. Every maximum clique of size $4$ is a Pasch configuration as shown in Fig. 4.

\subsection*{In dimension nine}

\begin{figure}[h]
\includegraphics[width=6cm]{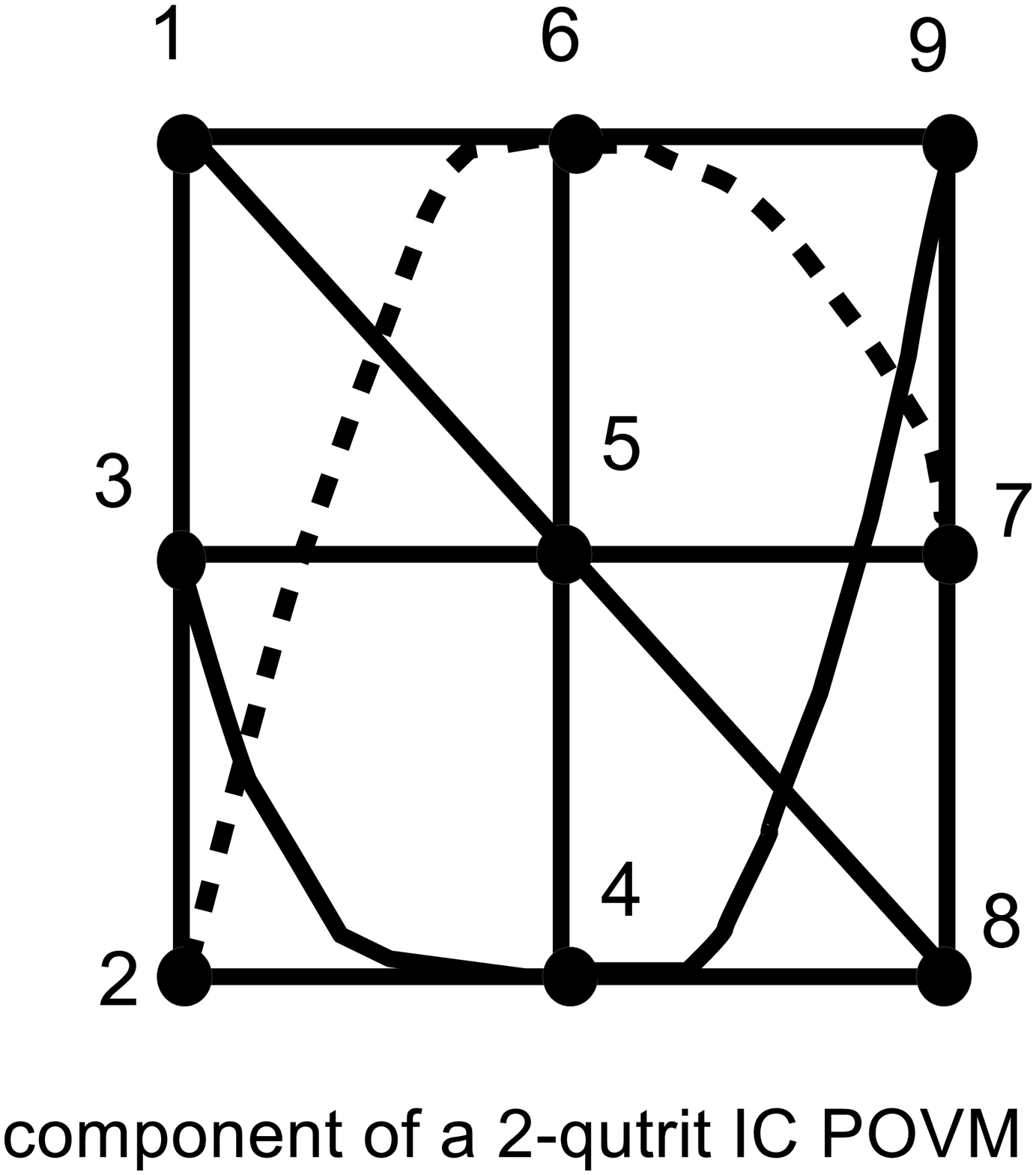}
\includegraphics[width=5cm]{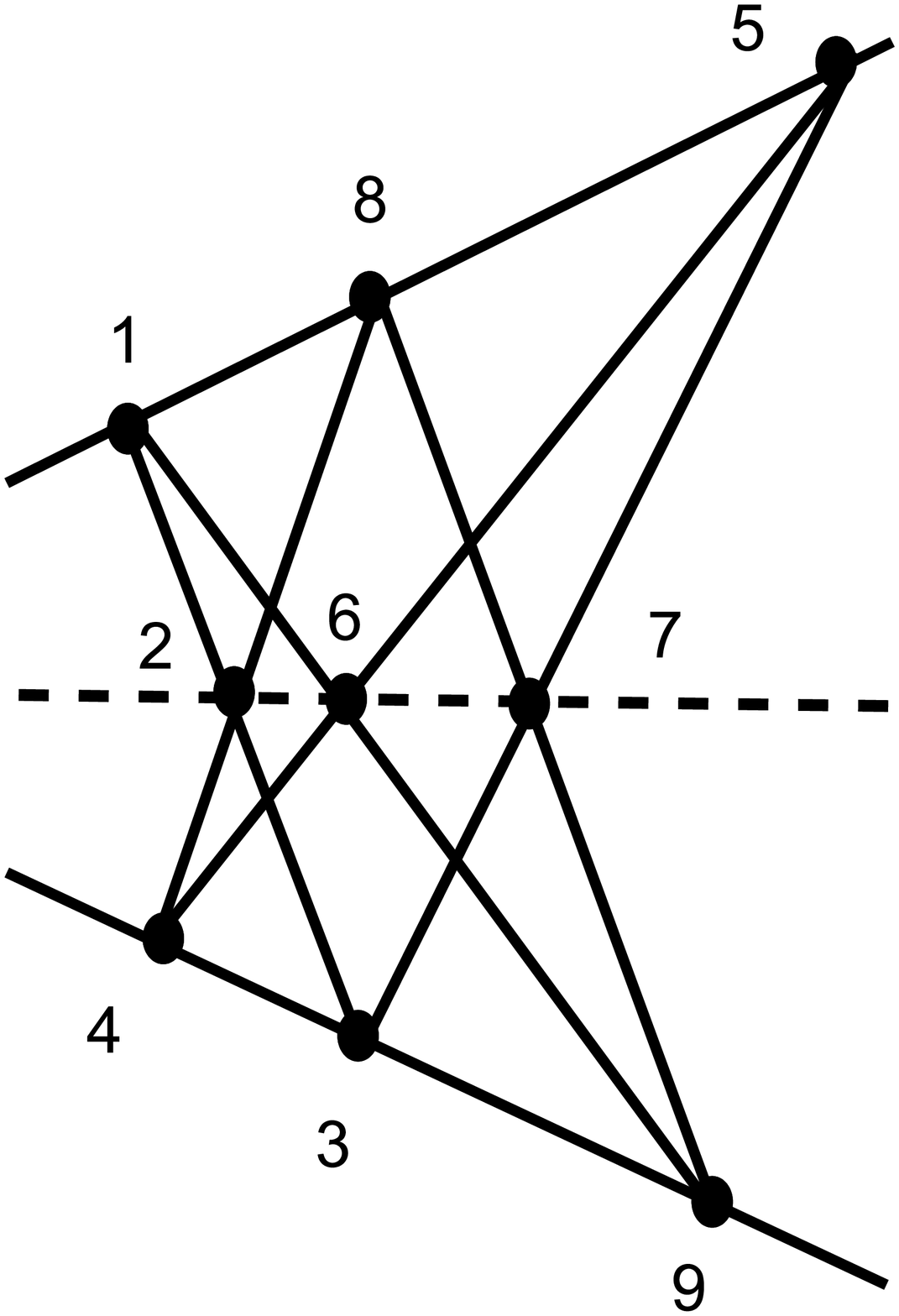}
\caption{Lines of one component of the two-qutrit IC-POVM built from the magic state $(1,1,0,0,0,0,-1,0,-1)$ alias the Pappus configuration (right figure). The points are labelled in terms of the two-qutrit operators $[1,2,3,4,5,6,7,8,9]=[I \otimes Z, I \otimes XZ, I \otimes (XZ^2)^2, Z \otimes I, Z \otimes X,
Z\otimes X^2, Z^2 \otimes Z^2, Z^2 \otimes (XZ)^2, Z^2 \otimes XZ^2]$, where $X$ and $Z$ are the qutrit shift and clock operators. The IC-POVM, as labelled, can be used to prove the Kochen-Specker theorem for two qutrits. This is related to the fact that the selected product of operators on a line is the identity matrix $\mathcal{I}$ except for the dotted line where it is $\omega_3 \mathcal{I}$ (see details in the text).
 }
\end{figure}

Let us consider a magic group isomorphic to $\mathbb{Z}_3^2  \rtimes \mathbb{Z}_4$ generated by two magic gates. One finds a few magic states such as $(1,1,0,0,0,0,-1,0,-1)$ that, not only can be used to generate a dichotomic IC-POVM with distinct pairwise products   $\left|\left \langle \psi_i \right|\psi_j \right\rangle|^2$ equal to $\frac{1}{4}$ or $\frac{1}{4^2}$, but also show a quite simple organization of triple products. Defining lines as triple of projectors with trace $\frac{1}{8}$, one gets a geometric configuration of type $[81_3]$ that split into nine disjoint copies of type $[9_3]$. One of the copies is shown in Fig. 5.

The configuration $[9_3]$ labelled by the operators of Fig. 5 may be used to provide an operator proof of the Kochen-Specker theorem with two qutrits. The proof is in the same spirit than the one derived for two or three qubits \cite{Planat2012} (see also \cite{Cabello2012} for two-qutrit contextuality). The vertices are projectors instead of just Hermitian operators. On one hand, every operator $O$ can be assigned a value  $\nu(O)$ which is an eigenvalue of $O$, that is $1$, $\omega_3$ or $\omega_3^2$ (with $\omega_3^3 =1$. Taking the product of eigenvalues over all operators on a line and over all nine lines, one gets $1$ since every assigned value occurs three times. 

On the other hand, the operators on a line in Fig. 5 do not necessarily commute but their product is $\mathcal{I}=I \otimes I$ , $\omega_3 \mathcal{I}$ or $\omega_3^* \mathcal{I}$, depending on the order of operators in the product. Taking the ordered triples $[1,6,9]$, $[9,7,8]$, $[2,4,8]$, $[1,3,2]$, $[8,5,1]$, $[3,5,7]$, $[3,4,9]$, $[4,5,6]$ and $[2,6,7]$, the triple product of these operators from left to right equals $\mathcal{I}$ except for the dotted line where it is $\omega_3 \mathcal{I}$.

Thus the product law $\nu(\Pi_{i=1}^9 O_i)=\Pi_{i=1}^9[\nu(O_i)]$ is violated. The left hand side equals $\omega_3$ while the right hand side equals $1$. No non-contextual hidden variable theory is able to reproduce these results. Since the lines are not defined by mutually commuting operators, it is not possible to arrive at a  proof of the two-qutrit Kochen-Specker based on vectors instead of operators. In this sense, the proof of contextuality is weaker that the one obtained for two or three qubits.

\subsection*{Higher dimensions}

The same method based on eigenstates of permutation matrices leads to IC-POVMs in dimensions higher than $9$. In the next subsection, we provide details about a $12$-dimensional IC-POVM covariant under the two-qubit/qutrit Pauli group because the associated triple products contain some geometrical structures as it was the case in lower dimensions.

\subsection*{In dimension twelve}

\begin{figure}[h]
\includegraphics[width=6cm]{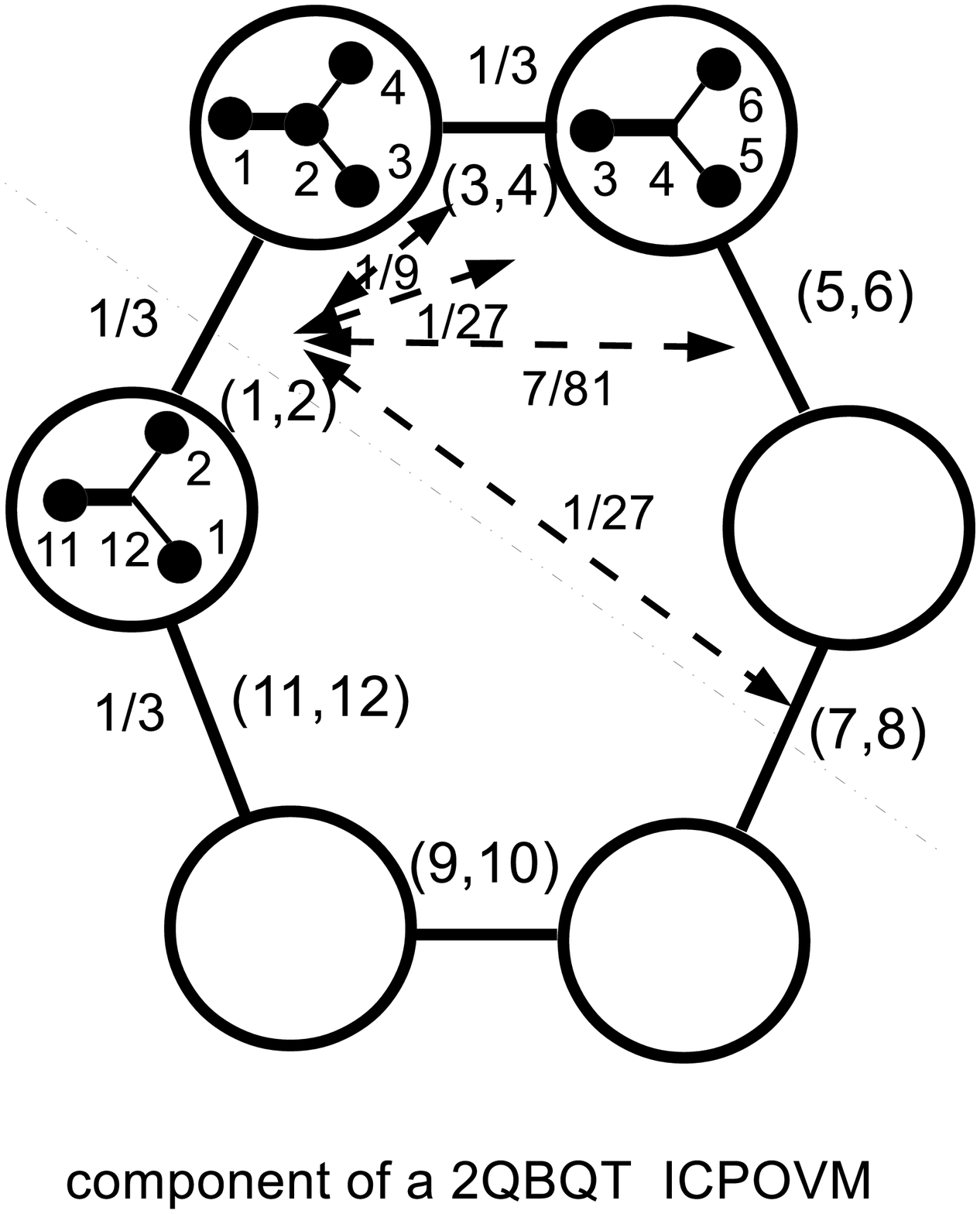}
\caption{Schematic of a $12$-projector component for the IC-POVM built from the magic state given in the text. Each circle contains two triples [e.g. $(1,2,3)$ and $(1,2,4)$ for the upper left circle]. The twelve projectors numbered $1$ to $12$ are passed by pairs from one circle to the other (as shown) so that empty circles are easily filled. The $4$ types of pair products $\frac{1}{3},\frac{1}{9},\frac{1}{27},\frac{7}{81}$ occurring are also shown.  
 }
\end{figure}

One can build an IC-POVM using a magic group isomorphic to $\mathbb{Z}_2^2 \rtimes(\mathbb{Z}_3^2 \rtimes \mathbb{Z}_2^2 )$. A magic state can be taken as 
$(0,1,\omega_6-1,\omega_6-1,1,1,\omega_6-1,-\omega_6,-\omega_6,0,-\omega_6,0)$ or the state obtained from it by permuting the entries $\omega_6-1$ and $-\omega_6$ and vice versa. The IC is obtained thanks to the action of the $2$QB-QT Pauli group on this state. The distinct trace products are multivalued with $8$ values. There is an interesting structure of the $144$ triple products whose trace is $-\frac{1}{27}$. There are organized into $12$ distinct configurations of the type shown in Fig. 6 (with only $4$ values of pair products occurring).

\small
\begin{table}[h]
\begin{center}
\begin{tabular}{|l|c|r|l|}
\hline 
\hline
dim& magic state  & $\left|\left \langle \psi_i \right|\psi_j \right\rangle|^2_{i \ne j}$ & Geometry\\
\hline
\hline
2  & $\left| T \right \rangle$ & $1/3$& tetrahedron \cite{Bravyi2004} \\
\hline
3  & $(0,1,\pm1 )$& $1/4$ & Hesse SIC \cite{Bengtsson2010} \\
\hline
4  & $(0,1,-\omega_6,\omega_6-1)$ & $\{1/3,1/3^2\}$ &  Mermin square$^*$ \\
\hline
5  & $(0,1,-1,-1,1)$& $1/4^2$ & Petersen graph \\
& $(0,1,i,-i,-1)$&  & \\
& $(0,1,1,1,1)$& $\{1/3^2,(2/3)^2\}$ & \\
\hline
\hline
6  & $(0,1,\omega_6-1,0,-\omega_6,0)$ & $\{1/3,1/3^2\}$ & Borromean rings \\
\hline
7  & $(1,-\omega_3-1,-\omega_3,\omega_3,\omega_3+1,-1,0)$ & $1/6^2$& unknown \\
\hline
8  & $(-1 \pm i,1,1,1,1,1,1,1)$& $1/9$ & Hoggar SIC \cite{Stacey2016}, $[63_3]^*$ \\
\hline
9  & $(1,1,0,0,0,0,-1,0,-1)$& $\{1/4,1/4^2\}$ &$[9_3]$ configuration$^*$ \\
\hline
12 & $(0,1,\omega_6-1,\omega_6-1,1,1,$ & $8$ values&Fig. 6\\
 & $\omega_6-1,-\omega_6,-\omega_6,0,-\omega_6,0)$ & & \\  
\hline
\hline
\end{tabular}
\caption{A summary of magic states and the corresponding signatures of IC-POVMs in dimensions $2$ to $12$. $^*$In dimensions $4$, $8$ and $9$, a proof of the two-qubit, two-qutrit and three-qubit Kochen-Specker theorem follows from the IC-POVM. For $d \ge 6$, the magic states leading to an IC (as distinguished) become rare. }
\end{center}
\end{table}
\normalsize
\newpage
\section{Summary and conclusion}

The main contribution of our work is the construction of asymmetric IC-POVMs built thanks to the action of the Pauli group on appropriate  permutation generated magic/fiducial states. A summary of the work is in Table 1.
It is remarkable that the same corpus of ideas may be used simultaneously for permutation groups, universal quantum computing, unambiguous quantum state recovery and also quantum contextuality. Further work may focus on extending the range of dimensions where IC-POVM's may be derived, relate the useful magic states to quantum error correction and state distillation \cite{Bravyi2004} and the Bayesian interpretation of quantum mechanics \cite{Fuchs2004}.

It is expected that this type of work will clarify the observed efficiency of quantum algorithms based on permutations \cite{Gedik2015} and the relation between contextuality and quantum computing \cite{Howard2012,Raussendorf2016}.









\section*{Funding}

 This work was supported by the French \lq\lq Investissements d'Avenir" program, project ISITE-BFC (contract ANR-15-IDEX-03).


\end{document}